\documentclass[iopart,twocolumn,superscriptaddress]{revtex4-2}
\usepackage{amssymb}
\usepackage{amsmath}
\usepackage{mathrsfs}
\usepackage{appendix}
\usepackage{graphicx}
\usepackage{float}
\usepackage[colorlinks=True,citecolor=blue,linkcolor=blue,allcolors=blue]{hyperref}
\usepackage[normalem]{ulem}
\usepackage{color}
\usepackage{bm}

\begin{document}
	
	\title{Characterizing generalized Floquet topological states in hybrid space-time dimensions}
	
	\author{Weiwei Zhu}
	\email{phyzhuw@ouc.edu.cn}
	\affiliation{College of Physics and Optoelectronic Engineering, Ocean University of China, Qingdao 266100, China}
	\affiliation{Key Laboratory for Optics Photoelectronics, Qingdao 266100, China}
	\affiliation{Engineering Research Center of Advanced Marine Physical Instruments and Equipment of MOE, Qingdao 266100, China}
	\author{Jian-Hua Jiang}
	\email{jhjiang3@ustc.edu.cn}
	\affiliation{School of Biomedical Engineering, Division of Life Sciences and Medicine, University of Science and Technology of China, Hefei 230026, China}
	\affiliation{Suzhou Institute for Advanced Research, University of Science and Technology of China, Suzhou, 215123, China}
	\affiliation{Department of Modern Physics, School of Physical Sciences, University of Science and Technology of China, Hefei, 230026, China}

	\begin{abstract}
		In spatiotemporally modulated systems, topological states exist not only in energy gaps but also in momentum gaps. Such unconventional topological states impose challenges on topological physics. The underlying models also make the conventional Hamiltonian descriptions complicated. Here, we propose to describe such systems with space- and time-direction transfer matrices which substantially simplify the underlying theory and give direct information on the topological properties of the quasienergy and quasimomentum gaps. In particular, we find that the space- and time-direction reflection phases can serve as signatures for distinguishing  various topological phases of the quasienergy and quasimomentum gaps. This approach directly reveals the topological properties of the band gap, avoiding the complexity in calculating bulk band topology in hybrid energy-moment space. By investigating two concrete models, we show that the method works well for both Hermitian and non-Hermitian systems. Furthermore, we uncover an unconventional topological state, called the anomalous Floquet quasimomentum gap, whose topological properties are invariant for different choices of the unit-cell center. This work advances the study of topological phenomena in hybrid space-time (energy-momentum) dimension that are attracting much interest due to the development of spatiotemporally modulated materials.
	\end{abstract}
	
	\maketitle
	
	\section{Introduction}\label{sec:Int}
	
	Topological properties of natural and artificial crystalline insulators with spatially periodic structures are usually described by topological (Bloch) band theory~\cite{RevModPhys.88.021004,RevModPhys.91.015006,Ma2019,zhu2022,Xue2022}. According to the bulk-edge correspondence, the topological edge states in an energy gap is determined by the topological invariant of all the bulk bands below the gap. The time analogue of crystals (i.e., temporally periodic systems, also called temporal crystals~\cite{PhysRevA.79.053821,1.4928659,014002}), can also support topological edge states but in momentum gaps~\cite{Lustig18,Giergiel_2019,Segal22,li2023}. Different from the spatial counterparts, temporal topological edge states are localized at time interfaces (e.g., a boundary in the temporal periodic structure). Topological invariants defined for bulk momentum bands can be used to describe the topological properties of such systems~\cite{Lustig18}.
	
	However, it is challenging to experimentally realize topological momentum gap in continuous system for the requirement of high modulation frequency and large modulation amplitude~\cite{Lustig18}. Discrete system, that can be realized in atomic system~\cite{science1174436,PhysRevLett.110.190601,Alberti_2014,PhysRevLett.121.070402,PhysRevLett.124.050502}, ionic system~\cite{PhysRevLett.103.090504,Matjeschk_2012,PhysRevLett.104.100503}, superconducting circuits~\cite{PhysRevX.7.031023,science.aaw1611,science.abg7812}, and optical systems~\cite{PhysRevLett.104.153602,Crespi2013,science.1193515,PhysRevLett.104.050502,PhysRevLett.106.180403,PhysRevLett.107.233902,PhysRevA.75.052310,PhysRevLett.110.263602,PhysRevLett.120.260501,Regensburger2012,Xiao2020,PhysRevLett.133.073803}, provide excellent platforms to study topological properties of momentum gap~\cite{ren2024,feis2024}. Besides, the discrete system is usually periodically modulated both in time and space, which provides the opportunity to studying topological properties of coexisting quasienergy and quasimomentum gaps in the same system~\cite{feis2024,zhu2024}. However, an efficient method to characterize the topological properties of such systems is still under developing. In this work, we propose to characterize the topological properties of quasienergy/quasimomentum gap for discrete models by directly considering the scattering properties of the band gap, avoiding the difficulty of calculating bulk band topology~\cite{PhysRevLett.106.057001,PhysRevB.85.165409,PhysRevB.93.075405,PhysRevX.4.021017,Xiao2015,PhysRevB.97.195307}.
	
	We design a general discrete model containing forward propagating channels, backward propagating channels and general scatterers, that can support quasienergy and/or quasimomentum gap by designing scatterers. Symmetrized space-direction transfer matrix and time-direction transfer matrix are derived to describe their scattering properties. The sign of obtained space-direction reflection phase and time-direction reflection phase can be used to distinguish different topological phases of quasienergy gap and quasimomentum gap~\cite{PhysRevX.4.021017,Lustig18}. This characterization method directly considers the properties of the band gap, avoiding the difficulty of calculating bulk band topology. 
	
	Specifically, we use such method to study two non-Hermitian models whose scatterers are parity-time symmetric dimers and their quasienergy band and quasimomentum band are intrinsically non-Hermitian. In the first model, pure quasienergy gap phases or pure quasimomentum gap phases are supported. We show their quasienergy gap and quasimomentum gap have anomalous Floquet topological nature~\cite{PhysRevLett.106.220402,PhysRevX.3.031005}, that topological properties are the same for unit cells with different inversion centers. In the second model, the system simultaneously supports quasienenrgy gap and quasimomentum gap at all parameters except for the phase transition point. Here the quasienergy gap still has anomalous Floquet topological nature, however the quasimomentum gap is more like a topological phase in static system whose topology can be tuned by changing the inversion center of unit cells.
	
	This paper is organized as follows. In Sec.~\ref{sec:The}, we introduce our model and outline its theoretical description including space-direction transfer matrix and time-direction transfer matrix. The definition of reflection phase is provided here. In Sec.~\ref{sec:ModI}, we use the method to study model I. We show different topological phases can be distinguished by the sign of reflection phase. We also show both quasienergy gap and quasimomentum gap of model I has anomalous Floquet topological nature. In Sec.~\ref{sec:ModII}, we use the method to study model II. Again, we show the method works well. Here for model I, the quasienergy gap has anomalous Floquet topological nature, while the quasimomentum gap is similar to a static topological gap. In Sec.~\ref{sec:Dif}, we discuss the difference between space-direction transfer matrix and time-direction transfer matrix. In Sec.~\ref{sec:Sum}, we conclude our study.
	
	\section{Theoretical Description}\label{sec:The}
	\subsection{Model and Transfer Matrix Description}
\begin{figure}
		\includegraphics[width=\linewidth]{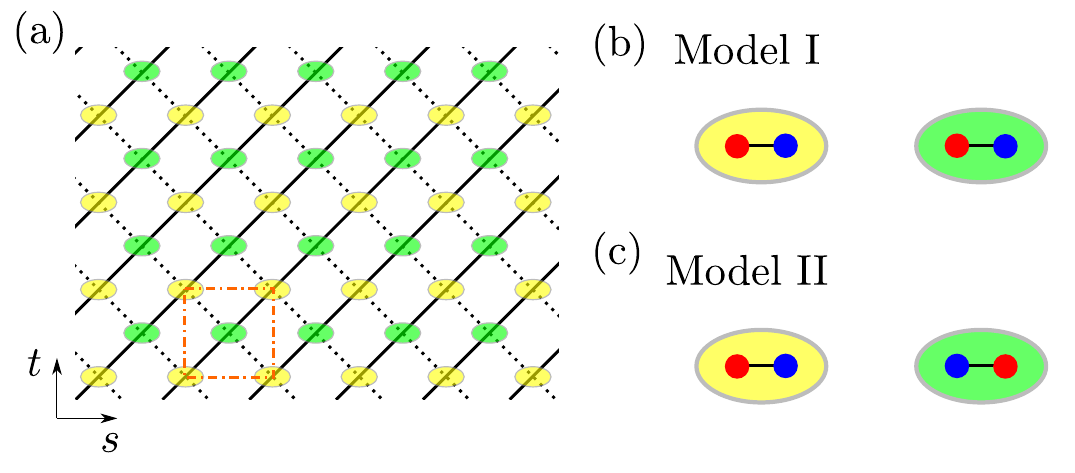}
		\caption{Model. (a) A general model which is composed forward propagating channels, backward propagating channels and general scatterers. Yellow (green) ellipse represents yellow (green) type scatterers with scattering matrix $U$ ($S$). (b) Scatterers for model I. (c) Scatterers for model II. Red (blue) sites represents gain (loss). Two sites are coupled with coupling strength $\theta$.}
		\label{model}
	\end{figure}

	The model we studied is shown in Fig.~\ref{model}(a), which is periodic both in space (horizontal) and time (vertical). The model is composed of forward propagating channels (solid line), backward propagating channels (dotted line) and scatterers (yellow ellipses and green ellipses). Each unit cell (dot-dashed box) contains two scatterers. Waves incident from forward propagating channels will be periodically scattered into backward propagating channels. The destructive interference of forward propagating waves and backward propagating waves give rise to forbidden Floquet-Bloch states in quasienergy gap or quasimomentum gap. Such model is quite general and can be realized in different systems, including atomic system~\cite{science1174436,PhysRevLett.110.190601,Alberti_2014,PhysRevLett.121.070402,PhysRevLett.124.050502}, ionic system~\cite{PhysRevLett.103.090504,Matjeschk_2012,PhysRevLett.104.100503}, superconducting circuits~\cite{PhysRevX.7.031023,science.aaw1611,science.abg7812} and optical systems~\cite{PhysRevLett.104.153602,Crespi2013,science.1193515,PhysRevLett.104.050502,PhysRevLett.106.180403,PhysRevLett.107.233902,PhysRevA.75.052310,PhysRevLett.110.263602,PhysRevLett.120.260501,Regensburger2012,Xiao2020,PhysRevLett.133.073803}. In this work, we will mainly study two specific models with scatterers shown in Figs.~\ref{model}(b) and \ref{model}(c). Before that, we first provide a general theory that can describe general scatterers.
	
	\begin{figure}
		\includegraphics[width=\linewidth]{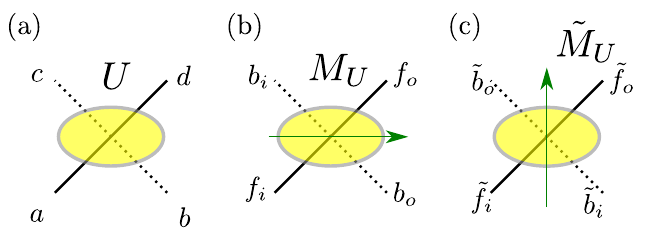}
		\caption{Schematic for a general scatterers $U$ with (a) scattering matrix representation, (b) space-direction transfer matrix representation, (c) time-direction transfer matrix representation.}
		\label{scattering}
	\end{figure}
	
	Fig.~\ref{scattering}(a) shows one scatterer, which connects $(c,d)^{T}$ and $(a,b)^{T}$ by
	
	\begin{equation}\label{eq1}
		\left(\begin{array}{c}
			c \\
			d
		\end{array}
		\right)=U\left(\begin{array}{c}
			a \\
			b
		\end{array}
		\right),
	\end{equation}
	$U$ called scattering matrix, is a two-by-two matrix, and can be simply obtained from matrix exponential of a two-by-two Hamiltonian. However, such description is not convenient to symmetrize the system, which is important to distinguish the topological properties of system by reflection phase~\cite{PhysRevX.4.021017}. A better description is the transfer matrix method. Fig.~\ref{scattering}(b) show the schematic of transfer matrix along space direction, which connect output signal in forward/backward channels $(f_o,b_o)^{T}$ and input signal in forward/backward channels $(f_i,b_i)^{T}$ by
	
	\begin{equation}\label{eq2}
		\left(\begin{array}{c}
			f_o \\
			b_o
		\end{array}
		\right)=M_U\left(\begin{array}{c}
			f_i \\
			b_i
		\end{array}
		\right),
	\end{equation}
	Eq.~\ref{eq2} can be obtained from Eq.~\ref{eq1} by a linear transformation with $f_o=d$, $b_o=b$, $f_i=a$ and $b_i=c$. The components of two-by-two matrix $M_U$ can be represented as $M_U(1,1)=(U_{12}U_{21}-U_{11}U_{22})/U_{12}$, $M_U(1,2)=U_{22}/U_{12}$, $M_U(2,1)=-U_{11}/U_{12}$ and $M_U(2,2)=1/U_{12}$. Similarly, we can describe the system by transfer matrix method in time direction. Fig.~\ref{scattering}(c) show the schematic of transfer matrix along time direction, which connects $(\tilde{f}_o,\tilde{b}_o)^{T}$ and $(\tilde{f}_i,\tilde{b}_i)^{T}$ by
	\begin{equation}\label{eq3}
		\left(\begin{array}{c}
			\tilde{f}_o \\
			\tilde{b}_o
		\end{array}
		\right)=\tilde{M}_U\left(\begin{array}{c}
			\tilde{f}_i \\
			\tilde{b}_i
		\end{array}
		\right),
	\end{equation}
	The components of two-by-two matrix $\tilde{M}_U$ can be represented as $\tilde{M}_U(1,1)=U_{21}$, $\tilde{M}_U(1,2)=U_{22}$, $\tilde{M}_U(2,1)=U_{11}$ and $\tilde{M}_U(2,2)=U_{12}$.
	
	Similarly, we can obtain space-direction transfer matrix and time-direction transfer matrix for green-type scatterers in Fig.~\ref{model}(a) from their scattering matrix $S$, and we represent them as $M_S$ and $\tilde{M}_S$, respectively.
	
	\subsection{Transfer Matrix for Unit Cell and Finite Structure}
	
	In this subsection, we derive the space direction transfer matrix and time direction transfer matrix for unit cell and finite structures. We also provide the definition of reflection phase, which is used to describe topological properties of quasienergy gap and quasimomentum gap latter.
	
	\begin{figure}
		\includegraphics[width=\linewidth]{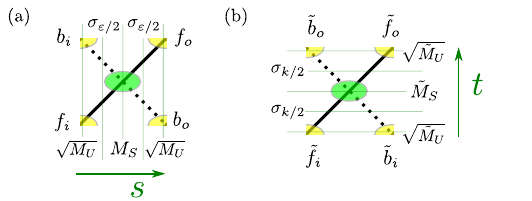}
		\caption{Schematic for transfer matrix of unit cell. (a) Space-direction transfer matrix. (b) Time-direction transfer matrix,}
		\label{trans}
	\end{figure}
	
	We first derive the space direction transfer matrix for unit cell. The schematic is shown in Fig.~\ref{trans}(a). In the derivation, we set periodic boundary conditions in the time direction. So the quasienergy $\varepsilon$ is good quantum number. In the space direction, the transfer matrix connects output signals $(f_o,b_o)^T$ and input signals $(f_i,b_i)^T$. From left to right, there are five processes as shown in Fig.~\ref{trans}(a). In the first one, the input signals are scattered by yellow type scatterer. To symmetrize the unit cell, here we only consider half yellow type scatterer. Such process can be described by $\sqrt{M_U}$. In the second process, the forward waves and backward waves accumulate transport phase $\varepsilon/2$ and $-\varepsilon/2$, respectively. Such process can be described by matrix $\sigma_{\varepsilon/2}$, which is defined as $m_{\varepsilon/2}=e^{i\varepsilon/2}(\sigma_0+\sigma_z)/2+e^{-i\varepsilon/2}(\sigma_0-\sigma_z)/2$ with $\sigma_0$ and $\sigma_z$ the identity matrix and Pauli matrix. In the third process, the signals are scattered by green type scatterer. Such process is described by $M_S$. The fourth and fifth processes are described by $m_{\varepsilon/2}$ and $\sqrt{M_U}$, respectively. In total, the transfer matrix for unit cell can be represented as
	
	\begin{equation}\label{eq4}
		M=\sqrt{M_U}.m_{\varepsilon/2}.M_S.m_{\varepsilon/2}.\sqrt{M_U}
	\end{equation}
	
	The quasimomentum band of the system $k(\varepsilon)$ can be obtained from
	
	\begin{equation}
		\det{\left[M-e^{ik(\varepsilon)}\right]}=0
	\end{equation}
	
	The transfer matrix for $N$ unit cells can be obtained from power function of matrix
	
	\begin{equation}
		\mathcal{M}=M^{N}
	\end{equation}
	We consider a physical process where a forward propagating wave is incident from left then there will be reflected backward propagating wave and transmitted forward propagating wave. Such process can be represented as
	\begin{equation}\label{eq7}
		\left(\begin{array}{c}
			t \\
			0
		\end{array}
		\right)=\mathcal{M}\left(\begin{array}{c}
			1 \\
			r
		\end{array}
		\right),
	\end{equation}
	From which we obtain reflection coefficient and transmission coefficient,
	\begin{equation}
		r=-\frac{\mathcal{M}_{21}}{\mathcal{M}_{22}},	t=\frac{1}{\mathcal{M}_{22}}
	\end{equation}
	The space-direction reflection phase can be obtained by
	\begin{equation}
		\phi_r=\arg(r)
	\end{equation}
	
	We then derive the time direction transfer matrix for unit cell. The schematic is shown in Fig.~\ref{trans}(b). Different from previous case, here we apply periodic boundary condition in space direction. So the quasimomentum $k$ is a good quantum number. In the time direction, the transfer matrix connects output signals $(\tilde{f}_o,\tilde{b}_o)^T$ and input signals $(\tilde{f}_i,\tilde{b}_i)^T$. Similarly, here there are also five processes as shown in Fig.~\ref{trans}(b) but from down to up. The five processed can be described by $\sqrt{\tilde{M}_U}$, $m_{k/2}$, $\tilde{M}_S$, $m_{k/2}$ and $\sqrt{\tilde{M}_U}$, respectively. $m_{k/2}$ is defined as $m_{k/2}=e^{ik/2}(\sigma_0+\sigma_z)/2+e^{-ik/2}(\sigma_0-\sigma_z)/2$. Then the time direction transfer matrix for unit cell is represented as
	
	\begin{equation}\label{eq10}
		\tilde{M}=\sqrt{\tilde{M}_U}.m_{k/2}.\tilde{M}_S.m_{k/2}.\sqrt{\tilde{M}_U}
	\end{equation}
	
	The quasienergy band of the system $\varepsilon(k)$ can be obtained from
	
	\begin{equation}
		\det{\left[\tilde{M}-e^{i\varepsilon(k)}\right]}=0
	\end{equation}
	
	The time direction transfer matrix for $N$ unit cells can be obtained from power function of matrix
	
	\begin{equation}
		\tilde{\mathcal{M}}=\tilde{M}^{N}
	\end{equation}
	
	We also consider a physical process where a forward propagating wave is incident from down. Here there is no reflected wave in the input sides for the wave is only propagating along the positive time direction. There are two output signals, one is in forward propagating channel and the other is in backward propagating channel. Such physical process can be represented as 
	\begin{equation}\label{eq13}
		\left(\begin{array}{c}
			\tilde{t}_f \\
			\tilde{t}_b
		\end{array}
		\right)=\tilde{\mathcal{M}}\left(\begin{array}{c}
			1 \\
			0
		\end{array}
		\right),
	\end{equation}
	
	Here we can define the transmission coefficient as
	
	\begin{equation}
		\tilde{t}\equiv \tilde{t}_f =\tilde{\mathcal{M}}_{11}
	\end{equation}
	
	Previous work has shown the phase difference between reflected wave and transmitted wave can be used to distinguish topological properties of photonic temporal crystal~\cite{Lustig18}. Here we define the time-direction reflection coefficient as 
	
	\begin{equation}
		\tilde{r}\equiv \frac{\tilde{t}_b}{\tilde{t}_f} =\frac{\tilde{\mathcal{M}}_{12}}{\tilde{\mathcal{M}}_{11}}
	\end{equation}
	The time reflection phase can be obtained by
	\begin{equation}
		\phi_{\tilde{r}}=\arg(\tilde{r})
	\end{equation}
	
	Previously works have shown the sign of reflection can be used to distinguish different topological phases of continuous model~\cite{PhysRevX.4.021017,Lustig18}. To prove the derived transfer matrix methods can be used to describe scattering properties of discrete system and distinguish different topological phases, in Appendix.~\ref{sec:App}, we use them to study some known or obvious phenomena. 
	
	\section{Results for Model I}\label{sec:ModI}
	
	In this section, we use the transfer matrix method to study the topological properties of model I, whose scatterers are shown in Fig.~\ref{model}(b). The red (blue) dot represents a lattice site with gain (loss). They are coupled with coupling strength $\theta$. The scattering matrix $U$ and $S$ can be represented as 
	\begin{equation}
		U=S=e^{iH_1}
	\end{equation}
	where $H_1=\theta\sigma_x+ig\sigma_z$. The model is equivalent to a time-dependent system with time-dependent Bloch Hamiltonian,
	\begin{eqnarray}
		H(k,t)=\left\{
		\begin{array}{cc}
			H_1&\ell T<t\leq \ell T+T/2\;\\
			\sigma_{k/2}H_1\sigma_{k/2}&\ell T+T/2<t\leq \ell T+T\;
		\end{array}
		\right.
	\end{eqnarray}
	
	\subsection{Phase Diagram}
	
\begin{figure}
		\includegraphics[width=\linewidth]{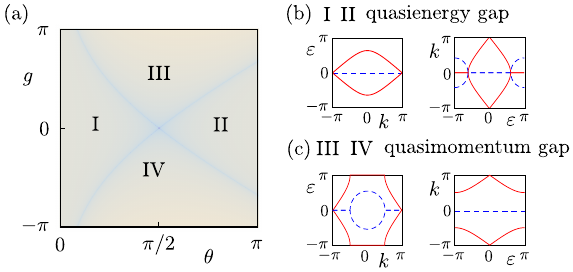}
		\caption{Phase diagram of model I. (a) Phase diagram. (b) An example of quasienergy band and quasimomentum band for phase I and phase II. (c) An example of quasienergy band and quasimomentum band for phase III and phase IV.}
		\label{modelI}
	\end{figure}

	Such model has been studied in earlier work~\cite{zhu2024} and shown to support quasienergy gap phase and quasimomentum gap phase (Fig.~\ref{modelI}). The merging of Floquet $\pi$ exceptional points plays an important role in the phase transition from quasienergy gap phase to quasimomentum gap phase. Here we study its topological properties in different phases.
	
	For convenience, here we show the phase diagram of model I in Fig.~\ref{modelI}(a), which has been shown in earlier work~\cite{zhu2024}. We notice two lines divide the parameter space into four parts, named I, II, III and IV. Phase I and II belong to quasienergy gap phase as one example shown in Fig.~\ref{modelI}(b). There is a quasienergy gap around quasienergy $\varepsilon=\pm\pi$. Phase III and IV belong to quasimomentum gap phase as one example shown in Fig.~\ref{modelI}(c). There is quasimomentum gap around quasimomentum $k=0$.
	
	\subsection{Space-direction Reflection Phase Used to Distinguish Quasienergy Gap}
	
\begin{figure}
		\includegraphics[width=\linewidth]{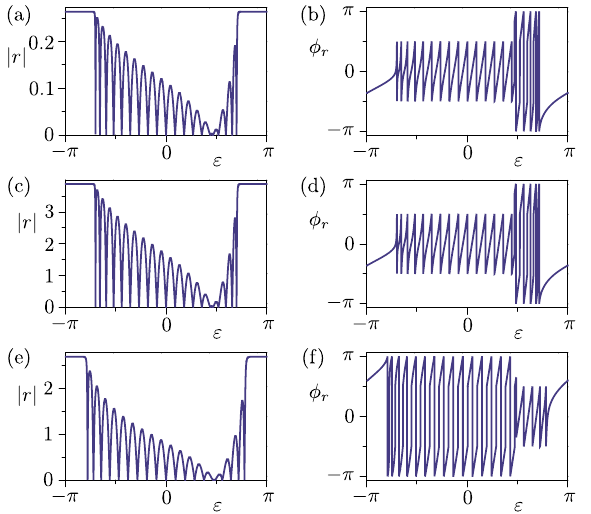}
		\caption{Space-direction reflection coefficients for model I. (a)(c)(e) Amplitude of the reflection coefficient. (b)(d)(f) Space-direction reflection phase. The parameters are $\theta=0.3\pi$, $g=0.8$ for (a)(b), $\theta=0.3\pi$, $g=-0.8$ for (c)(d), and $\theta=0.7\pi$, $g=0.8$ for (e)(f). The number of unit cells is chosen as $N=10$.}
		\label{sRP1}
	\end{figure}

	In this subsection, we use space-direction transfer matrix to describe the topological properties of quasienergy gap for model I. We first study the space-direction reflection coefficients for $\theta=0.3\pi$ and $g=0.8$,  which belong to phase I. The amplitude and phase of reflection coefficients are shown in Figs.~\ref{sRP1}(a) and \ref{sRP1}(b). We notice there is a band gap around quasienergy $\pm\pi$. However the amplitude of reflection coefficient is smaller than 1 for the non-Hermitian nature of the system. The reflection phase in full quasienergy gap is negative. We then exchange gain and loss ($\theta=0.3\pi$ and $g=-0.8$). The amplitude and phase of reflection coefficients are shown in Figs.~\ref{sRP1}(c) and \ref{sRP1}(d). In this case, the amplitude of reflection coefficients in quasinenrgy gap is changed to be larger than 1. However the reflection phase is still negative in the full quasienergy gap. It means exchanging gain and loss doesn't change the topological property of quasienergy gap.
	
	We then study the reflection coefficients for $\theta=0.7\pi$ and $g=0.8$,  which belong to phase II. The amplitude and phase of reflection coefficients are shown in Figs.~\ref{sRP1}(e) and \ref{sRP1}(f). We notice the amplitude of reflection coefficients in the quasienergy gap is larger than 1. Different from phase I, here the reflection phase is positive in full quasienergy band. 
	
\begin{figure}
		\includegraphics[width=0.6\linewidth]{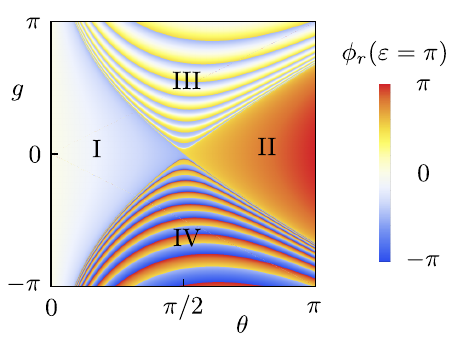}
		\caption{Space-direction reflection phase at $\varepsilon=\pm\pi$ for model I.}
		\label{sRP2}
	\end{figure}

	We then study the space-direction reflection phase at $\varepsilon=\pm\pi$ and consider the parameter space ($0\leqslant\theta\leqslant\pi$ and $-\pi\leqslant g\leqslant\pi$). The results are shown in Fig.~\ref{sRP2}. We notice in phase I, all the reflection phase is negative and in phase II, all the reflection phase is positive. So phase I and phase II are topologically different. Actually, phase I is topological trivial and phase II is topological nontrivial. Figs.~\ref{validity1}(a) and \ref{validity1}(b) are specific example of phase I and phase II without gain and loss ($g=0$). Besides, we notice the reflection phase at $\varepsilon=\pm\pi$ for phase III and phase IV undergo drastic change. The reason is that the quasienergy band of phase III and phase IV is gapless as shown in Fig.~\ref{modelI}(c). The drastic change of reflection phase comes from resonance mode around quasienergy $\pm\pi$.
	
	Usually, the topological properties of one-dimensional system can be changed by choosing unit cells with different inversion center. Here for model I, the topological properties of quasienergy $\pi$ gap doesn't change because the two scattering matrices $U$ and $S$ are the same and the transfer matrices for unit cell with different inversion center are the same. It means here the topological $\pi$ mode comes from anomalous Floquet topology~\cite{PhysRevLett.106.220402,PhysRevX.3.031005}.
	
	\subsection{Time-direction Reflection Phase Used to Distinguish Quasimomentum Gap}
	
	\begin{figure}
		\includegraphics[width=\linewidth]{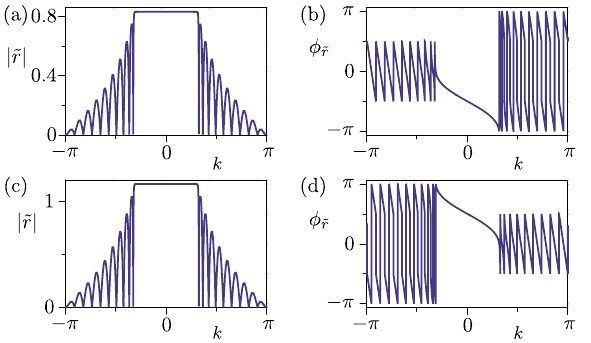}
		\caption{Time-direction reflection coefficient for model I. (a)(c) Amplitude of the reflection coefficient. (b)(d) Time-direction reflection phase. The parameters are $\theta=0.6\pi$, $g=0.9$ for (a)(b) and $\theta=0.6\pi$, $g=-0.9$ for (c)(d). The number of unit cells is chosen as $N=10$.}
		\label{tRP1}
	\end{figure}
	
	In this subsection, we use time-direction transfer matrix to describe the topological properties of quasimomentum gap for model I. We first study the time-direction reflection coefficients for $\theta=0.6\pi$ and $g=0.9$, which belong to phase III. The amplitude and phase of reflection coefficient are shown in Figs.~\ref{tRP1}(a) and \ref{tRP1}(b). From the amplitude, we notice there is a quasimomentum gap around $k=0$, which is consistent with the band structure in Fig.~\ref{modelI}(c). The amplitude of reflection coefficients in quasimomentum gap is smaller than 1, reflecting the non-Hermitian nature of the system. In full quasimomentum gap, the reflection phase is negative as shown in Fig.~\ref{tRP1}(b). We then exchange gain and loss ($\theta=0.6\pi$, $g=-0.9$). The amplitude and phase of reflection coefficient are shown in Figs.~\ref{tRP1}(c) and \ref{tRP1}(d). Here, the amplitude of reflection coefficients in quasimomentum gap is larger than 1. The reflection phase here is positive. Different sign of reflection phase means their topological properties are different.

	\begin{figure}
		\includegraphics[width=0.6\linewidth]{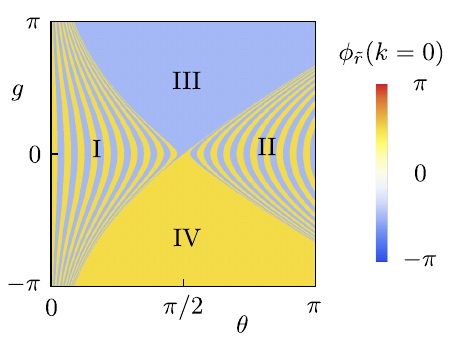}
		\caption{Time-direction reflection phase at $k=0$ for model I.}
		\label{tRP2}
	\end{figure}
	
	We then study the time-direction reflection phase at $k=0$ and consider the parameter space ($0\leqslant\theta\leqslant\pi$ and $-\pi\leqslant g\leqslant\pi$). The results are shown in Fig.~\ref{tRP2}. We notice the reflection phase is negative in phase III and positive in phase IV, showing they are topologically different. In phase I and phase II, the reflection phase dramatically change for the gapless quasimomentum band and resonance modes around $k=0$.
	
	Here when we choose a unit cell with different inversion center, the topological property of quasimomentum gap doesn't change for the same expression of $U$ and $S$. Although, the quasimomentum here is located at $k=0$, it has anomalous Floquet topological nature~\cite{PhysRevLett.106.220402,PhysRevX.3.031005}.
	
	\begin{figure}
		\includegraphics[width=\linewidth]{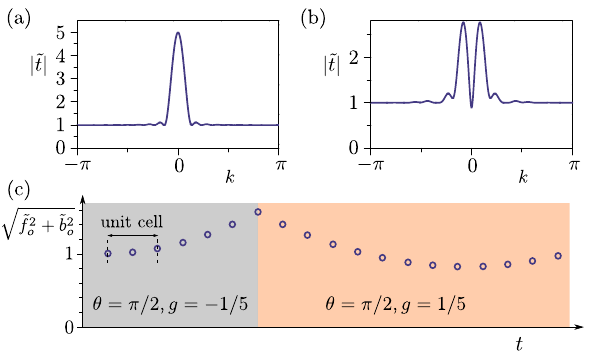}
		\caption{Time interface states for model I. (a) Amplitude of time-direction transmission coefficient for homogeneous structure with $\theta=\pi/2, g=-1/2$. The number of unit cells is chosen as $N=9$. (b) Amplitude of time-direction transmission coefficient for heterogeneous structure. Former three unit cells have parameters $\theta=\pi/2, g=-1/2$ and the later six unit cells have parameters $\theta=\pi/2, g=1/2$. (c) Field profile of time interface state at $k=0$.}
		\label{tInter1}
	\end{figure}
	
	Determined by bulk-boundary correspondence, there are topological interface states in the middle of heterogeneous structure composed of different topological phases. Here we consider a heterogeneous structure in time direction composed of $\theta=\pi/2, g=1/2$ in phase III and $\theta=\pi/2, g=-1/2$ in phase IV. As a comparison, we first study a homogeneous structure with parameters $\theta=\pi/2, g=-1/2$. The amplitude of time-direction transmission for homogeneous structure is shown in Fig.~\ref{tInter1}(a), where we find an amplification around $k=0$. The amplitude of time-direction transmission for heterogeneous structure is shown in Fig.~\ref{tInter1}(b), where we also find amplification around $k=0$ but with a downward peak at $k=0$. Such downward peak comes from the topological interface between phase III and phase IV. Fig.~\ref{tInter1}(c) show the field profile of topological interface state, where we notice the field first exponentially amplifies and then begins to decay at the interface, and after a period of time, it is amplified again.

	\section{Results for Model II}\label{sec:ModII}
	
	In this section, we use the transfer matrix method to study the topological properties of model II, whose scatterers are shown in Fig.~\ref{model}(c). Different from Model I, here the gain and loss sites of green type scatterer is exchanged. The scattering matrix $U$ and $S$ can be represented as 
	\begin{equation}
		U=e^{iH_1}, S=e^{iH_2}
	\end{equation}
	where $H_1=\theta\sigma_x+ig\sigma_z$, $H_2=\sigma_xH_1\sigma_x$. The model is equivalent to a time-dependent system with time-dependent Bloch Hamiltonian,
	\begin{eqnarray}
		H(k,t)=\left\{
		\begin{array}{cc}
			H_1&\ell T<t\leq \ell T+T/2\;\\
			\sigma_{k/2}H_2\sigma_{k/2}&\ell T+T/2<t\leq \ell T+T\;
		\end{array}
		\right.
	\end{eqnarray}
	
	\subsection{Phase Diagram}
\begin{figure}
		\includegraphics[width=\linewidth]{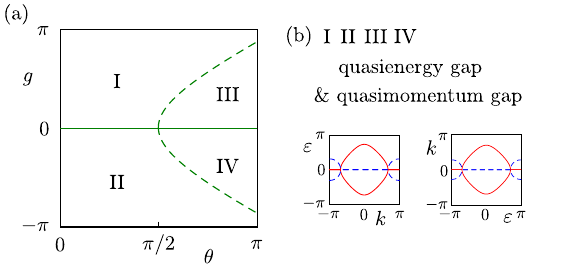}
		\caption{Phase diagram of model II. (a) Phase diagram. (b) An example of quasienergy band and quasimomentum band for phase I, phase II, phase III and phase IV.}
		\label{modelII}
	\end{figure}

	In this subsection, we first show phase diagram of the system. Fig.~\ref{modelII}(b) show one example of quasienergy band and quasimomentum band. We notice the system has quasienergy gap around $\varepsilon=\pm\pi$ and simultaneously has quasimomentum gap around $k=+\pm\pi$. Fig.~\ref{modelII}(a) show phase diagram of the system. The parameter space ($0\leqslant\theta\leqslant\pi$ and $-\pi\leqslant g\leqslant\pi$) is divided into four parts by green solid line and green dashed line. Green solid line happens at $g=0$, where the quasimomentum gap closed. Actually, at $g=0$ the system is Hermitian so the quasimomentum band is gapless. The green dashed line happens at $g=\pm\sqrt{\theta^2-(\pi/2)^2}$, where the quasienergy gap closed.
	
	Considering topological phase transition are usually accompanies by band gap close and reopening, we can conclude that phase I and phase II (or phase III and phase IV) have same kind of quasienergy gap, while phase I and phase III (or phase II and phase IV) have same kind of quasimomentum gap.
	
	\subsection{Space-direction Reflection Phase Used to Distinguish Quasienergy Gap}
	
\begin{figure}
		\includegraphics[width=\linewidth]{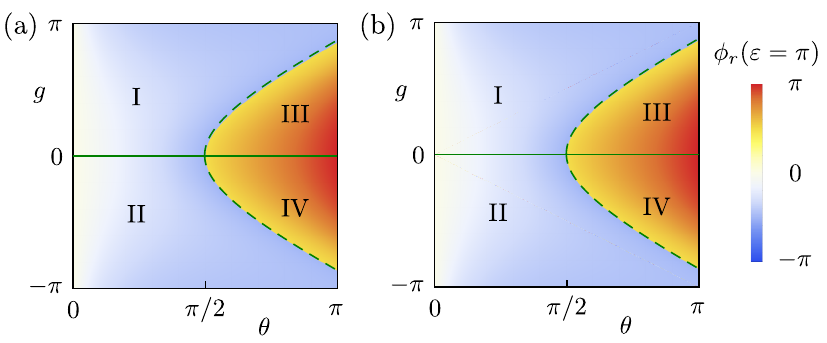}
		\caption{Space-direction reflection phase at $\varepsilon=\pm\pi$ for model II. (a) The center of unit cell is chosen at center of $S$. (b) The center of unit cell is chosen at center of $U$.}
		\label{sRP3}
	\end{figure}

	In this subsection, we use space-direction transfer matrix to describe the topological properties of quasienergy gap for model II. As before, here the reflection phase is also positive or negative in the full quasienergy gap around $\varepsilon=\pm\pi$. We don't show specific example anymore and directly study the space-direction reflection phase at $\varepsilon=\pm\pi$ for parameter space $0\leqslant\theta\leqslant\pi$ and $-\pi\leqslant g\leqslant\pi$. The result is shown in Fig.~\ref{sRP3}(a). We notice the reflection phase  is negative for phase I and phase II, and positive for for phase III and phase IV. Different sign of reflection phase means different topological properties. Actually, phase I and phase II are topologically trivial, while phase III and phase IV are topologically nontrivial. We also notice topological phase transition can be realized solely by gain and loss for $\theta>\pi/2$.
	
	Before, we chose center of $S$ as the inversion center of unit cell. Next we change the inversion center of unit cell to center of $U$.  The space-direction reflection phase at $\varepsilon=\pm\pi$ for parameter space $0\leqslant\theta\leqslant\pi$ and $-\pi\leqslant g\leqslant\pi$ is shown in Fig.~\ref{sRP3}(b). We notice the topological properties of the system doesn't change. It means the quasienenrgy gap around $\varepsilon=\pm\pi$ for model II has anomalous Floquet topological nature.
	
	\subsection{Time-direction Reflection Phase Used to Distinguish Quasimomentum Gap}
	
\begin{figure}
		\includegraphics[width=\linewidth]{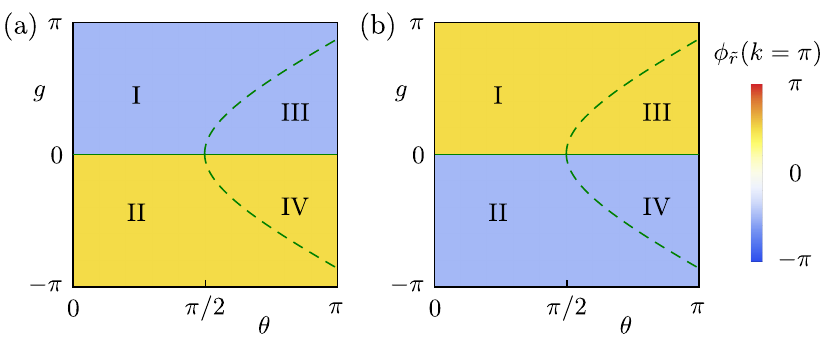}
		\caption{Time-direction reflection phase at $k=\pm\pi$ for model II. (a) The center of unit cell is chosen at center of $S$. (b) The center of unit cell is chosen at center of $U$.}
		\label{tRP3}
	\end{figure}

	In this subsection, we use time-direction transfer matrix to describe the topological properties of quasimomentum gap for model II. Also we don't show specific example anymore. We directly study the time-direction reflection phase at $k=\pm\pi$ for parameter space $0\leqslant\theta\leqslant\pi$ and $-\pi\leqslant g\leqslant\pi$. The result is shown in Fig.~\ref{tRP3}(a). We notice the quasimomentum gap of phase I and phase III have same topological properties with negative reflection phase, while phase II and phase III have same quasimomentum gap with positive reflection phase.
	
	Then we change the inversion center of unit cell to center of $U$, the time-direction reflection phase at $k=\pm\pi$ is shown in Fig.~\ref{tRP3}. We notice the topological properties of quasimomentum gap is changed. Here phase I and phase III have positive reflection phase, while phase II and phase IV have negative reflection phase. Here the topological property depending on the choice of inversion center of unit cell is usual like a topological state in static system, although the quasimomentum gap is located around $k=\pm\pi$. The topological property of such kind of quasimomentum gap is also studied in recent works.
	
	\begin{figure}
		\includegraphics[width=\linewidth]{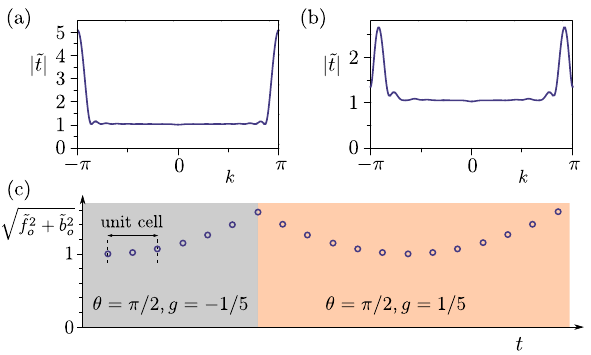}
		\caption{Time interface states for model II. (a) Amplitude of time-direction transmission coefficient for homogeneous structure with $\theta=\pi/2, g=-1/2$. The number of unit cells is chosen as $N=9$. (b) Amplitude of time-direction transmission coefficient for heterogeneous structure. Former three unit cells have parameters $\theta=\pi/2, g=-1/2$ and the later six unit cells have parameters $\theta=\pi/2, g=1/2$. (c) Field profile of time interface state at $k=\pm\pi$.}
		\label{tInter2}
	\end{figure}
	
	Here, we study a heterogeneous structure with different quasimomentum gap phase in time direction. The parameters we choose are $\theta=\pi/2$, $g=-1/5$ and $\theta=\pi/2$, $g=1/5$. As a comparison, we first study the time-direction transmission coefficients of a homogeneous structure with $\theta=\pi/2$, $g=-1/5$. The result is shown in Fig.~\ref{tInter2}(a), where we notice an amplification around $k=\pm\pi$. We then study the time-direction transmission coefficients of the heterogeneous structure. The result is shown in Fig.~\ref{tInter2}(b), where we find a downward peak at $k=\pm\pi$. Such peak comes from the topological interface state between different topological phases. Fig.~\ref{tInter2}(c) shows the field profile of topological interface state in time. As before, we notice the field first exponentially amplifies and then begins to decay at the interface, and after a period of time, it is amplified again.

	\section{Difference between Space-direction Transfer Matrix and Time-direction Transfer Matrix}\label{sec:Dif}
	
	In this section, we discuss the difference between space-direction transfer matrix and time-direction transfer matrix. As shown in Eq.~\ref{eq4} and Eq.~\ref{eq10}, the time-direction transfer matrix has similar formula with the space-direction transfer matrix. However their physical meaning is totally different, which contributes to different phenomenon, $e.g.$ the field in quasimomentum gap is amplified as shown in Fig.~\ref{tInter1}(a) and Fig.~\ref{tInter2}(b), while the field in quasienergy gap is spatially decaying. The space-direction transfer matrix describes a physical process described by Eq.~\ref{eq7}, which a steady state of the system. One forward propagating wave is incident from the input side of finite structure, after sufficient interference, reflected wave and transmitted wave leave the system from input side and output side, respectively. The reflected wave, which is a backward propagating wave at input side, is determined by the incident wave and physical properties of finite structure. The time-direction transfer matrix describes a physical process described by Eq.~\ref{eq13}. One forward propagating wave is incident from the input side, the reflected wave and transmitted wave leave the system both from output side. Differently, here the backward propagating wave at input side is unrelated with incident wave. Such difference between space-direction transfer matrix and time-direction transfer matrix intrinsically comes from the causality that the field at a certain moment is determined by past experiences and has nothing to do with the future.
	
	\section{Conclusion and discussions}\label{sec:Sum}
	
	We proposed to characterize topological phases of quasienergy gap and quasimomentum gap by space- and time-direction transfer matrices. The method directly considers the properties of the band gap, avoiding the difficulty of calculating bulk band topology. We design two concrete models, one supports pure quasienergy gap or pure quasimomentum gap, the other simultaneously supports both the quasienergy gap and quasimomentum gap. We show their topological properties can be characterized by the sign of the reflection phases in the space- and time-directions. We also find a new topological phase called the anomalous Floquet quasimomentum gap in the first model, whose topological property is independent of the choice of the inversion center of the unit cell.

	\section*{Acknowledgement}
	W. Z. acknowledges support from the Start up Funding from the Ocean University of China and the National Natural Science Foundation of China (Grants No.~12404499). J. H. J. thanks supports from the National Key Research and Development Program of China (Grant No. 2022YFA1404400), the ``Hundred Talents Program" of the Chinese Academy of Sciences, and National Natural Science Foundation of China (Grant No. 12125504), and the Priority Academic Program Development of Jiangsu Higher Education Institutions.
	
	\appendix
	\section{Validity of Transfer Matrix Description}\label{sec:App}
	To demonstrate the validity of transfer matrix method derived, here we use them to study some known or obvious phenomena. We consider the scatterers as shown in Fig.~\ref{model}(b) without gain and loss. Then the scattering matrix $U$ and $S$ can be expressed as 
	
	\begin{equation}
		U=S=e^{i\theta\sigma_x}
	\end{equation}
	where $\sigma_x$, $\sigma_y$ and $\sigma_z$ are Pauli matrices. Here the model is equivalent to a time-dependent system, whose time-dependent Bloch Hamiltonian is
	
	\begin{eqnarray}
		H(k,t)=\left\{
		\begin{array}{cc}
			\theta\sigma_x&\ell T<t\leq \ell T+T/2\;\\
			\theta\sigma_{k/2}\sigma_x\sigma_{k/2}&\ell T+T/2<t\leq \ell T+T\;
		\end{array}
		\right.
	\end{eqnarray}
	where $\ell\in\mathbb{Z}$ and $\sigma_{k/2}$ is defined as $\sigma_{k/2}=e^{ik/2}\sigma_++e^{-ik/2}\sigma_-$ with $\sigma_\pm=(\sigma_x\pm i\sigma_y)/2$. $T$ is the time period and we set $T=2$. 
	
	\begin{figure}
		\includegraphics[width=\linewidth]{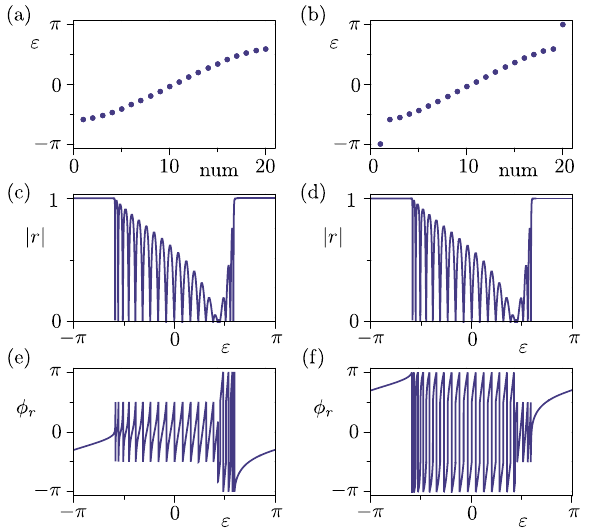}
		\caption{(a)(b) Spectrum of a finite structure. (c)(d) Amplitude of space-direction reflection coefficient. (e)(f) Space-direction reflection phase. The parameters are $\theta=0.3\pi$ for (a)(c)(e) and $\theta=0.7\pi$ for (b)(d)(f). The number of unit cells is chosen as $N=10$.}
		\label{validity1}
	\end{figure}
	
	Such model is well known to be normal quasienenrgy $\pi$ gap insulator with $0<\theta<\pi/2$ and support topological $\pi$ modes with $\pi/2<\theta<\pi$. Figs.~\ref{validity1}(a) and \ref{validity1}(b) show the quasienergy spectrum of finite structure for $\theta=0.3\pi$ and $\theta=0.7\pi$, where we notice there are topological $\pi$ modes for $\theta=0.7\pi$. So Fig.~\ref{validity1}(a) and Fig.~\ref{validity1}(b) belong to different topological phases. We then use the space direction transfer matrix method to describe them. Figs.~\ref{validity1}(c) and \ref{validity1}(e) show the amplitude and phase of reflection coefficient for $\theta=0.3\pi$. We notice around quasienengy $\pm\pi$, the amplitude of is 1, meaning quasienergy gap around quasienergy $\pm\pi$. In full band gap, the reflection phase is negative as shown in Fig.~\ref{validity1}(e). Figs.~\ref{validity1}(d) and \ref{validity1}(f) show the amplitude and phase of reflection coefficient for $\theta=0.7\pi$. We notice there is also a band gap around quasienergy $\pm\pi$. However, the reflection phase is positive in full band gap as shown in Fig.~\ref{validity1}(f). Early works have shown the sign of reflection phase can be used to distinguish different topological phases in one dimensional system with inversion symmetry. Here we see the transfer matrix method we derived works well and can be used to distinguish different topological phases.
	
	\begin{figure}
		\includegraphics[width=\linewidth]{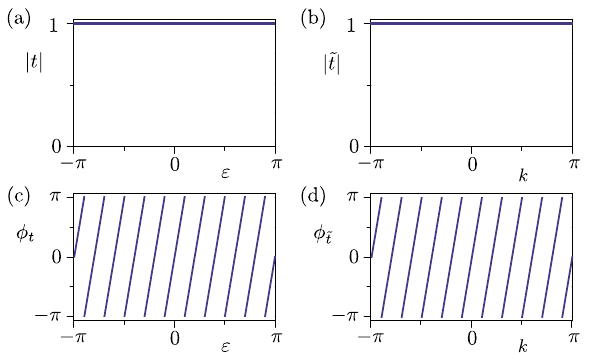}
		\caption{(a) Amplitude of space-direction transmission coefficient. (b) Amplitude of time-direction transmission coefficient. (c) Space-direction transmission phase. (d) Time-direction transmission phase. The parameter in calculation is $\theta=\pi/2$, the number of unit cells is chosen as $N=10$.}
		\label{validity2}
	\end{figure}
	
	We then use the transfer matrix method to describe a specific case for $\theta=\pi/2$, where obviously an incident forward propagating wave keep in forward propagating channels and will not be scattered into back propagating channels. Figs.~\ref{validity2}(a) and \ref{validity2}(c) show the amplitude and phase of space direction transmission coefficients. We find the amplitude is unit for all quasienergies and the phase is linearly dependent on quasienergies $\varepsilon$. We observe a similar phenomena for the time direction transmission coefficients as shown in Figs.~\ref{validity2}(b) and \ref{validity2}(d). These results are consistent with the inference that an incident forward propagating wave keep in forward propagating channels. It confirms the validity of transfer matrix we derived.

\end{document}